\documentclass[aps,nofootinbib,onecolumn,citeautoscript]{revtex4}
\usepackage{amsmath,amssymb}
\usepackage{hyperref}
\usepackage{enumerate}
\usepackage{graphicx}

\graphicspath{{/home/kapil/My_data/kapilpaper/arXiv/Robustness/}}
\usepackage{graphicx}
\usepackage{float}
\date{\today}
\begin{document}
\title{Robustness of Greenberger\textendash Horne\textendash Zeilinger and W states against Dzyaloshinshkii-Moriya interaction}

\author{Kapil K. Sharma$^\ast$ and S. N. Pandey$^\dagger$  \\
\textit{$^\ast$Department of Electrical Engineering, Indian Institute of Technology Bombay, \\ Mumbai 400076, India} \\
E-mail: $^\ast$iitbkapil@gmail.com\\ 
\textit{$^\dagger$Department of Physics, Motilal Nehru National Institute of Technology, \\ Allahabad-211004, India} \\
Email: $^\dagger$snp@mnnit.ac.in}

\begin{abstract}
In this article the robustness of tripartite W and Greenberger\textendash Horne\textendash Zeilinger (GHZ) states is investigated  against Dzyaloshinskii-Moriya (i.e. DM) interaction. We consider a closed system of three qubits and an environmental qubit.  The environmental qubit interacts with any one of the three qubits through DM interaction. The tripartite system is initially prepared in W and GHZ states respectively. The composite four qubits system evolve with unitary dynamics. We detach the environmental qubit by tracing out from four qubits and profound impact of DM interaction is studied on the initial entanglement of the system. As a result we find that the bipartite partitions of W states suffer from entanglement sudden death (i.e. ESD), while tripartite entanglement does not. On the other hand bipartite partitions and tripartite entanglement in GHZ states does not feel any influence of DM interaction. So, we find that GHZ states have robust character than W states. In this work we consider generalised W and GHZ states and three $\pi$ is used as an entanglement measure. This study can be useful in quantum information processing where unwanted DM interaction takes place.
\end{abstract}
\maketitle

\section{Introduction}
\label{intro}
Entanglement is a fascinating concept in quantum mechanics which has no classical description \cite{EPR1935,Neilsen2000}. It is the phenomenon which gives thrust to new technological development in quantum information processing. The roots of entanglement come from the debate of Einstein and Schr{\"o}dinger in the Era of 1935 \cite{sc1,sc2}. This concept attracted the attention of quantum community when it's first application as teleportation is proposed in a seminal paper by C. H. Bennett et al. in 1993 \cite{bent}, which has experimental verification also in many quantum systems \cite{ex}. Up to till date, entanglement has applications in almost all the streams such as in information security, game theory, image processing, quantum secrete sharing, superdense coding and quantum machine learning  and others \cite{AkEkert1991,qs,qi,qc,qm1,qm2,qm3}. The problem starts with entanglement when entangled quantum systems interact with the environment, which may destroy it. If entanglement suddenly dies for a finite time then this phenomenon is called entanglement sudden death (i.e. ESD), which is investigated by Yu and Eberly \cite{YuEberly2004,YuEberly2009}. The quantum systems which suffer from ESD for a particular time of interval may not be useful for quantum applications. So, quantum community is always interested to search such systems which can persist long entanglement. By keeping in view of application part of quantum systems, it is important to investigate the situations for ESD in bipartite to multipartite cases. There are many literature available on ESD in bipartite cases with Dzyaloshinskii-Moriya (i.e. DM) interaction \cite{z1,z2,z3,k1,k2,k3,k4,k5}. Here in the present paper we investigate this phenomenon in bipartite and tripartite case with DM interaction. DM interaction has an important role in quantum information processing, which takes place between two qubits connected through a ligand in a crystal \cite{d1,d2,d3}. If all the three entities (i.e. two qubits and ligand) are collinear then DM interaction becomes zero. Here we express, the possibility of  unwanted DM interaction that may arise when a foreign qubit adjust in the vacancy defect present in any crystal. Recently it has been found that this interaction also plays an important role to produce skyrimons \cite{sk1,sk2,sk3,sk4}, which are expected key candidates for future data storage devices. The impact of this interaction has been studied in varieties of quantum spin chains even with thermal conditions \cite{th1,th2}. 

The motivation of this study come from our previous work done in bipartite cases \cite{k1,k2,k3,k4,k5}. In this present work we consider a tripartite closed system of three qubits and an environmental qubit. The environmental qubit establish DM interaction with any one of the qubit of the tripartite system and disturb the entanglement. The tripartite system is prepared initially in generalized W and Greenberger\textendash Horne\textendash Zeilinger (GHZ) states respectively and environmental qubit is prepared in pure state \cite{gh,w1,w2}. Both the states have significant role in  quantum information and always
production of these in different physical systems, knowing their properties and dynamics are the subject of study in many situations \cite{o1,o2,o3,o4,o5,o6}. In past few years, the tripartite ESD is studied in pure and mixed W and GHZ states in different paradigms \cite{te1,te2}. To study the dynamics in tripartite system considered in the paper, we trace out the environmental qubit and  investigate the disturbance of DM interaction and the state of environmental qubit on initial entanglement of the system. We find, the state of environmental qubit does not disturb the entanglement in the system irrespective weather the system is prepared in W and GHZ states. It is only DM interaction which has periodic influence on entanglement. It periodically amplifies the entanglement and sometimes kills it in many quantum states. DM interaction produces ESD only in bipartite partitions of W states, while tripartite entanglement does not suffer from ESD. On the contrary, neither the bipartite partitions nor the tripartite entanglement in GHZ states face ESD, because the present study reveals that these states are not fragile with respect to (w.r.t.) the environmental DM interaction. The outlines of the paper is as follows. 

In Sect. 2, we discuss the Hamiltonian of the system, time evolution, W and GHZ states. Sect. 3, presents the discussion on three $\pi$ measure for entanglement. Further Sect. 4 is divided in two parts, in first part we consider the case 1 with W states and in second part case 2 is considered with GHZ states. Lastly the conclusion is presented in Sect. 5. 
\section{Hamiltonian, time evolution, W and GHZ states.}
\label{sec:1}
In this study the goal is to investigate the disturbance of DM interaction by taking an environmental qubit on tripartite system. Let consider tripartite system formed by qubits A, B and C, which takes place in $2\otimes 2\otimes 2$ dimensional Hilbert space. The environmental qubit (named D) interacts with any one qubit (assumed qubit C) of the tripartite system through DM interaction. Here we consider this interaction in z direction. The pictorial representation of the system is shown in Fig. 1. 
So, the Hamiltonian is framed as given below
\begin{center}
\begin{eqnarray}
H_{z}=D_{z}.(\sigma_C^X \otimes \sigma_D^Y-\sigma_C^Y \otimes \sigma_D^X),    \label{eq:Hz}   
\end{eqnarray}
\end{center}
where $\sigma_C^X$ and $\sigma_C^Y$ are  Pauli matrices for qubit (C) and $\sigma_D^X$, $\sigma_D^Y$ are Pauli matrices of qubit (D). Further $D_{z}$ is the DM interaction strength in z direction. The above Hamiltonian is easily diagonalizable by using the method of eigendecomposition. Further the unitary time evolution operator can be computed as 
\begin{eqnarray}
U(t)=e^{-i H t}. \label{eq:20}
\end{eqnarray}
The action of this operator will be used to obtain the time evolution of the tripartite system. By using this operator the time evolution density matrix of the composite system formed by four qubits A, B, C and D can be obtained as below
\begin{equation}
\rho(t)=U(t)\rho(0)U(t)^{\dagger}, \label{te}
\end{equation}
where $\rho(0)=\rho_{s}(0)\otimes \rho_{e}(0)$. The factor $\rho_{s}(0)$ corresponds to the initial state of tripartite system and $\rho_{e}(0)$ is the initial density matrix of the environmental qubit. The environmental qubit is prepared in pure state given as 
\begin{figure*}
        \centering
                \includegraphics[width=0.3\textwidth]{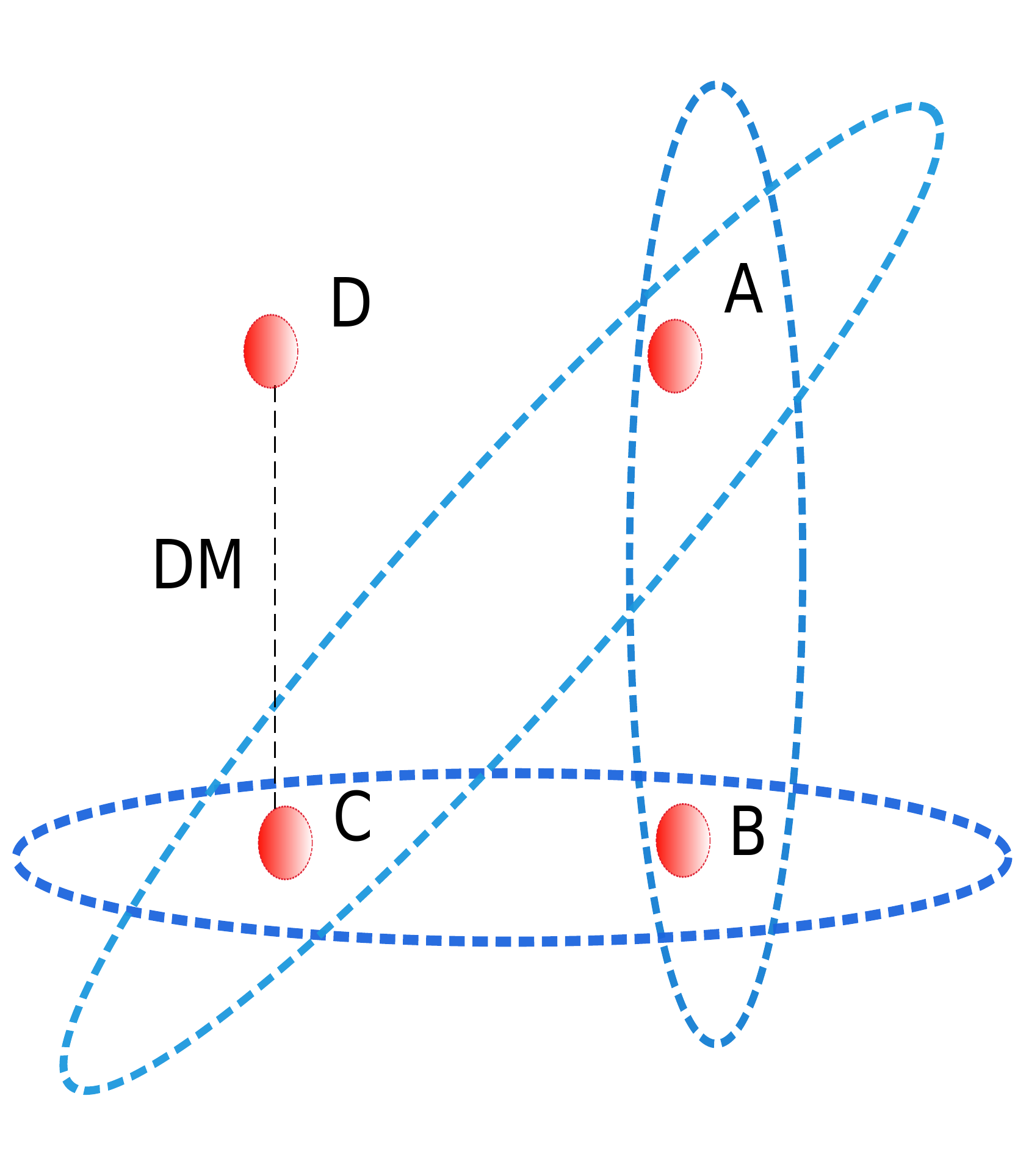}
                \caption{The tripartite system consisted of three entangled qubits A, B and C. DM interaction takes place between qubits D and C. Where D is the envrionemental qubit.}\label{sys}
\end{figure*}
\begin{equation}
|\psi\rangle=c_{0}|0\rangle+c_{1}|1\rangle,
\end{equation}
with 
\begin{equation}
|c_{0}|^{2}+|c_{1}|^{2}=1. \label{pa}
\end{equation}
Here we mention that the tripartite system initially prepared in generalised W and GHZ states respectively. The generalised W states are given below 
\begin{equation}
|\psi_{w}\rangle=w_{0}|001\rangle+w_{1}|010\rangle+w_{2}|100\rangle, \label{ws}
\end{equation}
with
\begin{equation}
|w_{0}|^{2}+|w_{1}|^{2}+|w_{2}|^{2}=1. \label{ncw}
\end{equation}
These states are mapped to bipartite entangled states when any one of the qubit is lost and carry pairwise entanglement. The GHZ states have opposite property. The generalised GHZ states for three qubits can be written as
\begin{equation}
|\psi_{G}\rangle=g_{0}|000\rangle+g_{1}|111\rangle, \label{gs}
\end{equation}
with
\begin{equation}
|g_{0}|^{2}+|g_{1}|^{2}=1. \label{ncg}
\end{equation}
These states mapped to bipartite disentangled states when any one qubit is lost, so do not carry pairwise entanglement and have only three way entanglement. Here first we focus on tripartite entanglement in W states and later for entanglement in it's bipartite partitions. To study the tripartite entanglement, one needs mathematical tool to measure it. Here we use three $\pi$ measure for pure states \cite{tpi,mn}, which is given in the next section which have connection with negativity \cite{n1,n2,n3,n4,n5}. Negativity is a good measure of entanglement in bipartite systems.
\section{The Entanglement measure (Three $\pi$)} 
There are many measures for entanglement but few of these do not fit to measure entanglement in some tripartite quantum states. For example three tangle in terms of concurrence $C$ is able to measure the tripartite entanglement in GHZ state but fails in W states and in many other pure states. 
The concurrence $C$ in a density matrix 
$\rho$ is given by 
\begin{equation}
C(\rho)=max\{0,\lambda_{1}-\lambda_{2}-\lambda_{3}-\lambda_{4}\},
\end{equation}
where $\lambda_{i}$, $i=1,2,3,4$, are the square roots of the eigenvalues in decreasing order of $\rho \rho^{s}$, $\rho^{s}$ is the spin flip density matrix given as \\
\begin{equation}
\rho^{s}=(\sigma^{y}\otimes \sigma^{y}) \rho^{*}( \sigma^{y}\otimes \sigma^{y}), 
\end{equation}
here $\rho^{*}$ is the complex conjugate of the density matrix $\rho$ and $\sigma^{y}$ is the Pauli Y matrix. The matrix $\rho \rho^{s}$ is used to calculate the concurrence. This concurrence in tripartite systems is used to calculate the tangle, which is a tripartite entanglement measure. The three tangle inspired by CKW inequality \cite{ckw} in terms of concurrence is defined as
\begin{equation}
\tau_{ABC}=C_{A|BC}^{2}-C_{AB}^{2}-C_{AC}^{2},\label{ttan}
\end{equation}
where $C_{A|BC}$ is the concurrence between qubit A and qubits B and C together. $C_{AB}$ is  the concurrence between  qubits A and B. $C_{AC}$ is the concurrence between qubits A and C.
Here the calculated expression of the three tangle for a generalized pure three qubit states $|\psi\rangle=\sum_{i,j,k}x_{i,j,k}|ijk\rangle$, is given as
\begin{equation}
\tau_{ABC}(|\psi\rangle)=4|4a_{3}-2a_{2}+a_{1}|,\label{tt}
\end{equation}
Where 
\begin{eqnarray}
a_{3}=x_{000}x_{110}x_{101}x_{011}+x_{111}x_{001}x_{010}x_{100},\\
a_{2}=x_{000}x_{111}x_{011}x_{100}+x_{000}x_{111}x_{101}x_{010}\\ \nonumber
+x_{000}x_{111}x_{110}x_{001}+x_{011}x_{100}x_{101}x_{010}\\ \nonumber
+x_{011}x_{100}x_{110}x_{001}+x_{101}x_{010}x_{110}x_{001}, \\ 
a_{1}=x_{000}^{2}x_{111}^{2}+x_{011}^{2}x_{110}^{2}+x_{010}^{2}x_{101}^{2}+x_{100}^{2}x_{011}^{2}.
\end{eqnarray}
Calculating the quantity $\tau_{ABC}$ by using the  (\ref{tt}) for the W states given in Eq. (\ref{ws}), we get $\tau_{ABC}(|\psi_{w}\rangle)=0$. While for GHZ states given in Eq. (\ref{gs}) this quantity is obtained as $\tau_{ABC}(|\psi_{G}\rangle)=4|g_{0}g_{1}|^{2}\textgreater 0$. For the special case of GHZ state $(|\psi_{G}\rangle=\frac{1}{\sqrt{2}}|000\rangle+\frac{1}{\sqrt{2}}|111\rangle)$, it is obtained as  $\tau_{ABC}(|\psi_{G}\rangle)=1\textgreater 0$. So we recall here that three tangle becomes zero for W states and does not seem a good measure for tripartite entanglement in W states. So for our work we use an entanglement measure called three $\pi$, which is greater than zero for W and GHZ states and capture the feature of measurement of entanglement in both the states. 
\begin{figure*}
        \centering
                \includegraphics[width=0.8\textwidth]{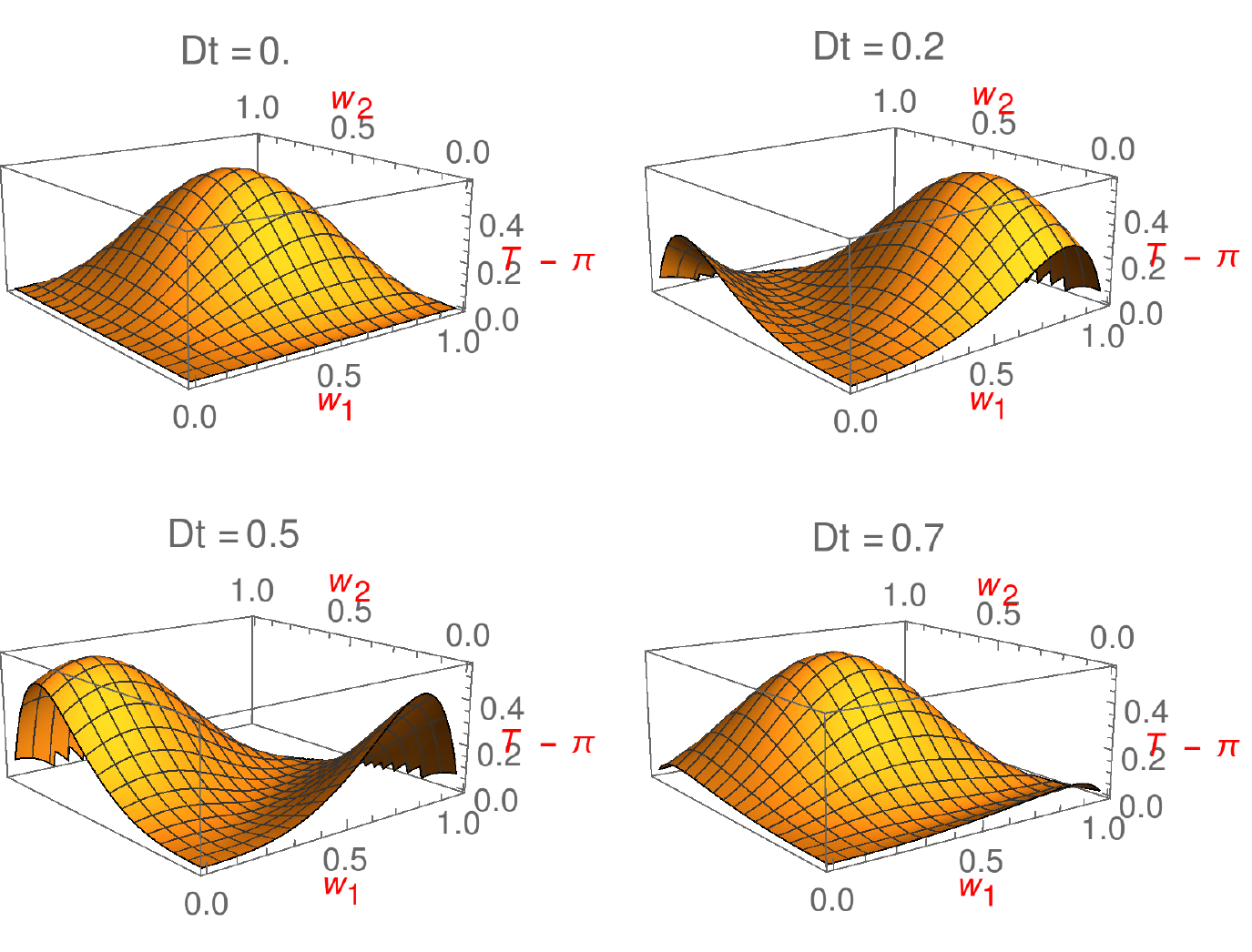}
                \caption{Plot of three $\pi$ (i.e. T-$\pi$) with different values of parameter $Dt$ (i.e. $Dt=0.0, 0.2, 0.5, 0.7$).}\label{2}
\end{figure*}
The three $\pi$ measure is defined as the average of three kinds of residual entanglement by concentrating on different nodal qubits in tripartite system. The idea of residual entanglement come from CKW like monogamy inequality for negativity \cite{ckw}. Let we concentrate on qubit A (i.e. nodal qubit A) in tripartite system shown in Fig. \ref{sys}, than the CKW monogamy inequality with negativity is given as 
\begin{equation}
N_{A|BC}^{2}\geq N_{AB}^{2}+N_{AC}^{2}. \label{abc}
\end{equation}   
where the negativity is defined as ``\textit{The absolute sum of all negative eigenvalues of the density matrix after the perse partial transpose
is taken}".\\
Further, evaluation of Eq. (\ref{abc}) with equality sign leads as
\begin{equation}
\pi_{A}=N_{A|BC}^{2}-N_{AB}^{2}-N_{AC}^{2}. \label{a}
\end{equation}
The term $\pi_{A}$ is called residual entanglement in tripartite system, when nodal qubit is considered as A. Similarly when the nodal qubit is considered B and C respectively, the terms of residual entanglement are given as 
\begin{eqnarray}
\pi_{B}=N_{B|AC}^{2}-N_{BA}^{2}-N_{BC}^{2}. \label{b} \\
\pi_{C}=N_{C|AB}^{2}-N_{CA}^{2}-N_{CB}^{2}. \label{c}
\end{eqnarray}
\begin{figure*}
        \centering
                \includegraphics[width=0.8\textwidth]{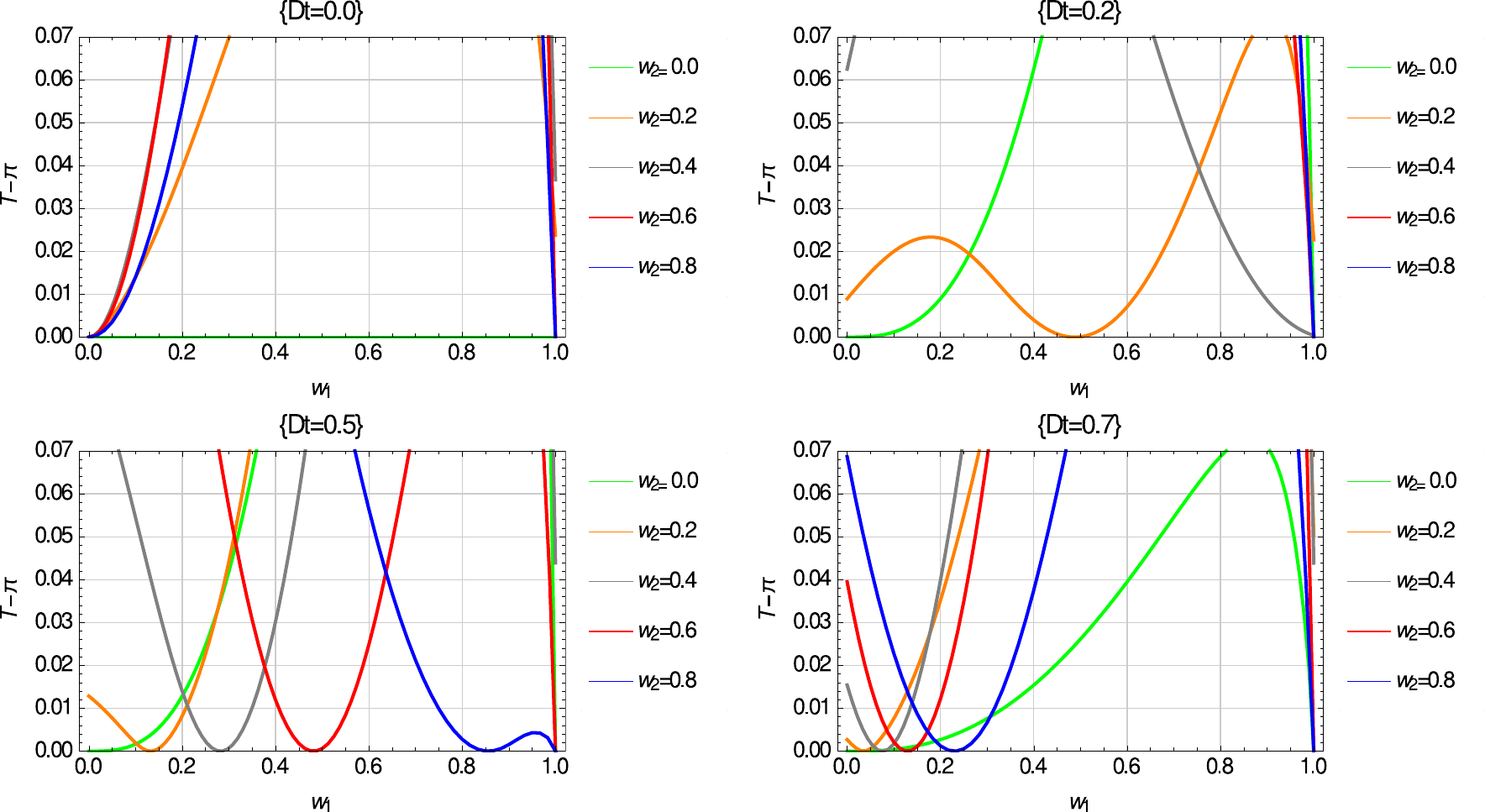}
                \caption{Two dimensionl plots corrosponding to Fig. $2$, with different values of parameter $Dt$ and $w_{2}$.}\label{3}
\end{figure*}
In the above equations, $N_{A|BC}$ is the negativity measure of entanglement of qubit A with qubits B and C together. In other words, the quantity express that how much qubit  A is entangled with qubits B and C together. Similar meanings are carried out by the terms  $N_{B|AC}$ and $N_{C|AB}$ when nodal qubits are considered B and C respectively. Further $N_{AB}$ is the bipartite negativity between partition AB of the tripartite system. And similarly, the meaning exists for $N_{BC}$ and $N_{CA}$ for bipartite partitions BC and CA. From the symmetry of entanglement the following factors are equal in Eqs. (\ref{a}), (\ref{b}) and (\ref{c}).
\begin{eqnarray}
N_{AB}=N_{BA}. \\
N_{BC}=N_{CB}. \\
N_{AC}=N_{CA}.
\end{eqnarray}
In general $(\pi_{A}\neq\pi_{B}\neq \pi_{C})$, which implies that residual entanglement varies under the permutations of nodal qubits A, B and C in tripartite system. Finally the three $\pi$  measure is given as the average of the terms $\pi_{A}$, $\pi_{B}$ and $\pi_{C}$. This average does not change under permutations of qubits A, B and C, it is given below.
\begin{equation}
\frac{1}{3}(\pi_{A}+\pi_{B}+\pi_{C}). \label{avp}
\end{equation}
\begin{figure*}
        \centering
                \includegraphics[width=0.8\textwidth]{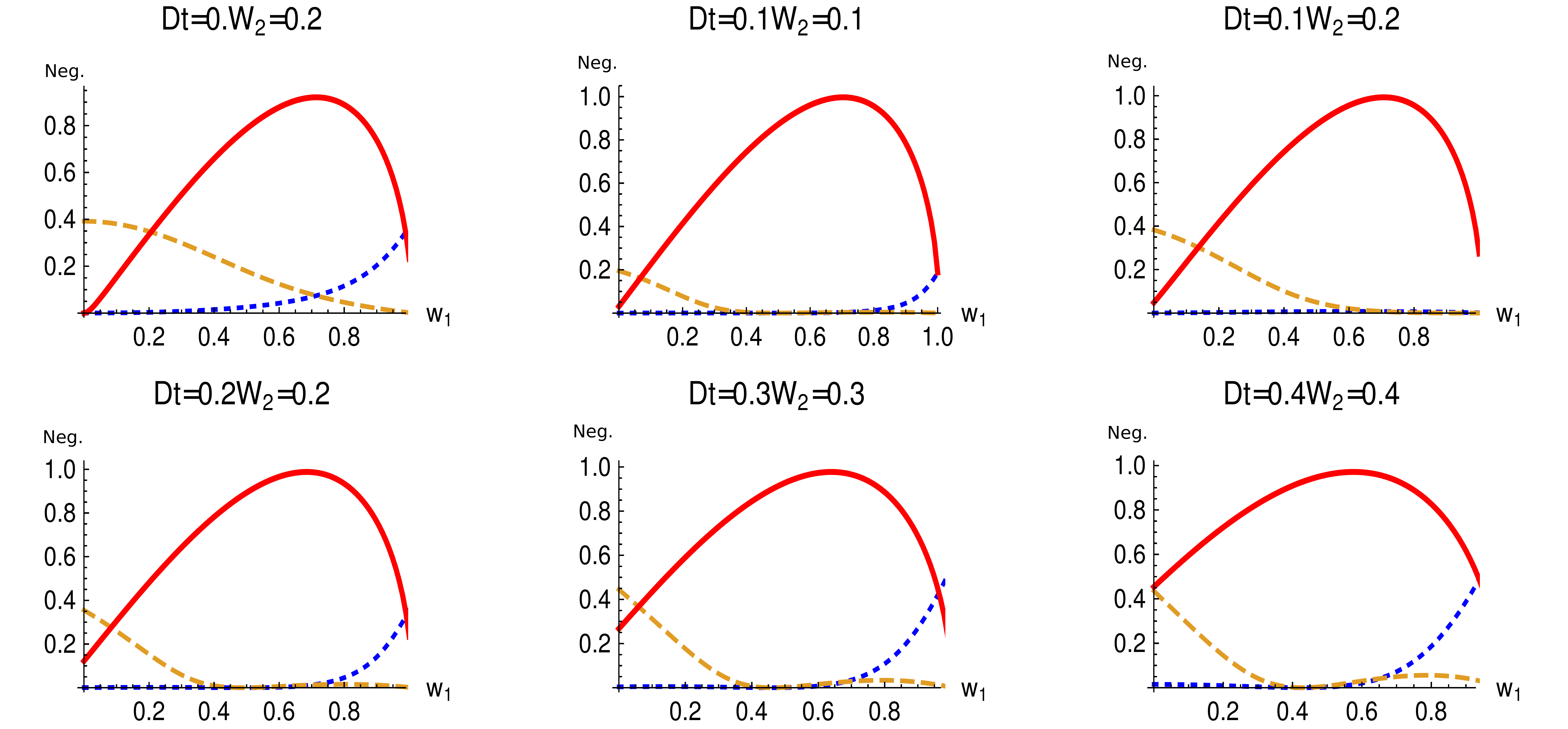}
                \caption{Plots of bipartite entanglement in generalised W states with different values of parameters $Dt$ and $w_{2}$. The quantities $N_{AB}$, $N_{AC}$ and $N_{BC}$ are represented by dotted blue, dashed orange and thick red graphs respectively.}\label{4}
\end{figure*}
We have calculated three $\pi$ for generalised W states given in Eq.(\ref{ws}) which is obtained as
\begin{equation}
\frac{4}{3}[w_{2}^{2}\sqrt{w_{2}^{4}+4w_{1}^{2}w_{0}^{2}}+w_{1}^{2}
\sqrt{w_{1}^{4}+4w_{2}^{2}w_{0}^{2}}+w_{0}^{2}
\sqrt{w_{0}^{4}
+4w_{2}^{2}w_{1}^{2}}-w_{2}^{4}-w_{1}^{4}-w_{0}^{4}].
\end{equation}
This factor achieve the values greater than zero for W states and has the ability to capture the measurement of entanglement in the states. For the special case of W states $(|\psi_{w}\rangle=\frac{1}{\sqrt{3}}|001\rangle+\frac{1}{\sqrt{3}}|010\rangle+\frac{1}{\sqrt{3}}|100\rangle)$,  this factor acquire the value $0.549364>0$. Here we mention that three $\pi$ for GHZ states achieve the value $1$. So it is concluded that the three $\pi$ is greater than zero for both the pure three qubit W and GHZ states. This measure also satisfies the properties which must be adopted by a good entanglement measure. These properties are given below.
\begin{itemize}
\item[a)] It must be zero for fully separable or bi-separable states and must acquire non zero value for fully entangled states.
\item[b)] It must be non increasing under local operations and classical communication (LOCC).
\item[c)] It must be invariant under local unitary operations (i.e. LU). 
\end{itemize}
Here we recall the quantity- three $\pi$ given in Eq. (\ref{avp}) which is invariant under the permutations of nodal qubits A, B and C and also invariant under the three simultaneous LU transformations. This measure satisfy all the properties given in $a)$, $b)$ and $c)$. For lucid mathematical proof of the above properties one may refer to Refs. \cite{mn,proof}. 
\section{Reducing the system and reckoning three $\pi$ measure}
In this section we find out the reduced density matrices for tripartite system. In Sect.\ref{sec:1}, we framed the Hamiltonian of the system and time evolution density matrix for four qubits A, B, C and D as given in Eq. (\ref{te}). The equation involve the initial density matrix of the system as $\rho(0)$. Here we mention that first we prepare the system in generalised W states and later in GHZ states. So we divide the study into two cases. 
\subsection{Case 1: Initial preparation with W states} 
We first prepare the system in generalised W states as given in Eq. (\ref{ws}) and find out the corresponding density matrix $\rho_{w}(0)$. The initial density matrix of the system is obtained as $\rho(0)=\rho_{w}(0)\otimes \rho_{e}(0)$, this matrix is plugged in Eq. (\ref{te}). Further we detach the environmental qubit by taking the partial trace operation over the basis of environmental qubit D and reduced density matrix $Rd_{ABC}$ is obtained as \\
\begin{equation}
Rd_{ABC}=[a_{ij}]_{8\times 8}.
\end{equation}
The matrix involve the terms $(a_{32},a_{23},a_{52},a_{25},a_{53},a_{35},a_{22},a_{33},a_{55})$ and rest of the terms are zero. The non zero terms of the matrix are given below.
\begin{center}
$\begin{array}{cccccccc}
a_{23}=a_{32}=(|c_{0}|^{2}+|c_{1}|^{2}) \sqrt{1-w_1^2-w_2^2}(w_2 \sin [2 \text{Dt}]+w_1 \cos [2 \text{Dt}]), \\ \\
a_{25}=a_{52}=-(|c_{0}|^{2}+|c_{1}|^{2}) \sqrt{1-w_{1}^{2}-w_{2}^{2}}(w_{1} \sin [2 \text{Dt}]-w_{2} \cos [2 \text{Dt}], \\ \\
a_{35}=a_{53}=\frac{1}{2}(|c_{0}|^{2}+|c_{1}|^{2})(w_{2}^{2}-w_{1}^{2})(\sin [4 \text{Dt}]+2 w_{1} w_{2} \cos [4 \text{Dt}]),\\ \\
a_{22}=-(|c_{0}|^{2}+|c_{1}|^{2})(-1+w_{1}^{2}+w_{2}^{2}), \\ \\
a_{33}=(|c_{0}|^{2}+|c_{1}|^{2})(w_{1} \sin [2 \text{Dt}]+w_{2} \cos [2 \text{Dt}])^{2},\\ \\
a_{55}=(|c_{0}|^{2}+|c_{1}|^{2})(w_{1} \sin [2 \text{Dt}]-w_{2} \cos [2 \text{Dt}])^{2}.
\end{array}$
\end{center}
\begin{figure*}
        \centering
                \includegraphics[width=0.8\textwidth]{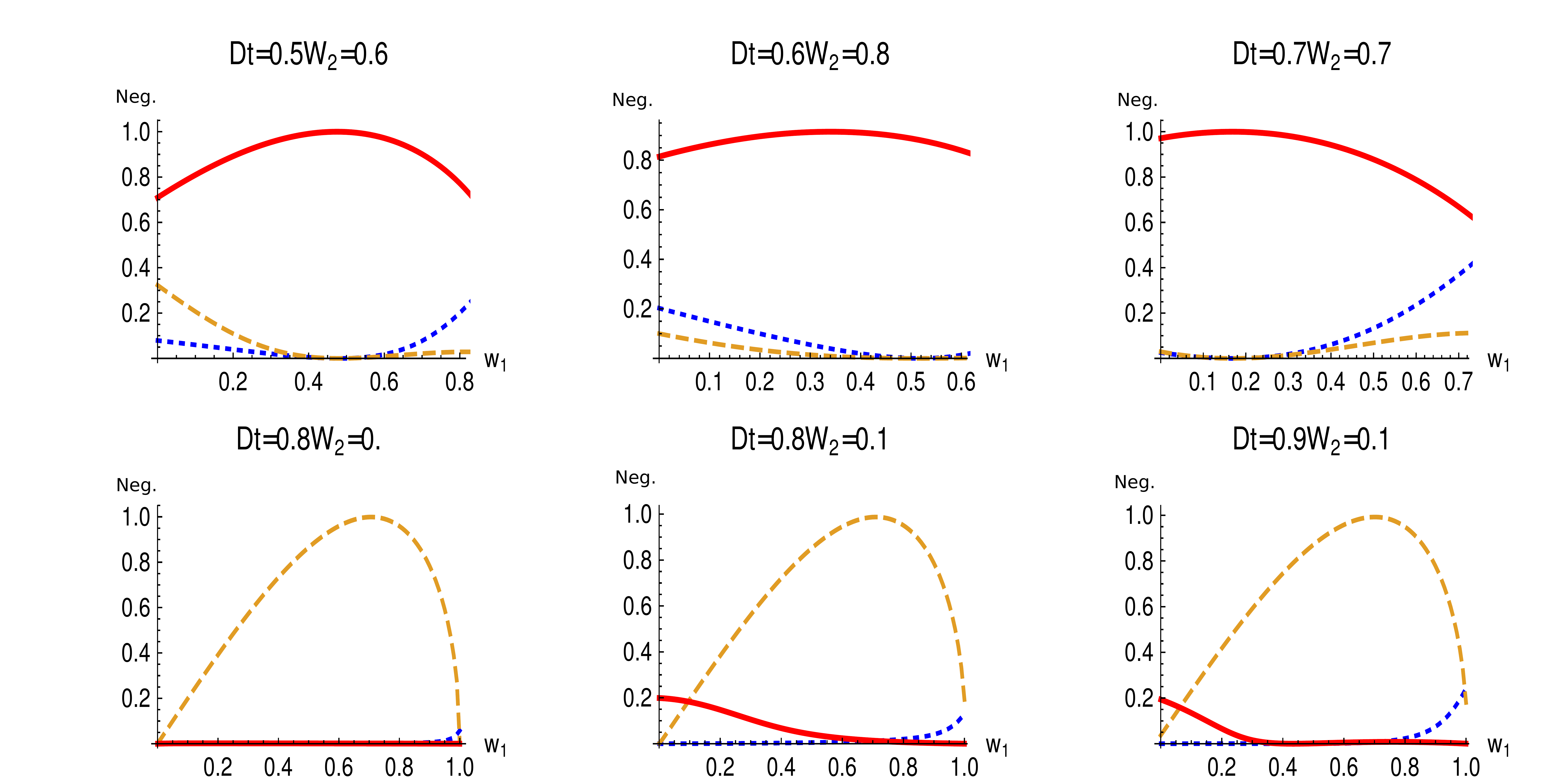}
                \caption{Plots of bipartite entanglement in generalised W states with different values of parameter $Dt$ and $w_{2}$. The quantities $N_{AB}$, $N_{AC}$ and $N_{BC}$ are represented with Dotted blue, Dashed orange and thick red graph respectively.}\label{5}
\end{figure*}
While calculating reduced density matrix, every term of the matrix involve the factor $|c_{0}|^{2}+|c_{1}|^{2}$. So by applying the normalization condition given in Eq. (\ref{pa}), the reduced density matrix is free from probability amplitude terms (i.e. $c_{0}$, $c_{1}$) of environmental qubit D. Hence the state of environmental qubit D does not have any impact on entanglement carried out by tripartite system. While on the other hand the parameter of DM interaction strength $(D_{z}=D)$ is involved in the reduced density matrix. The probability amplitudes of W states $(w_{0}$, $w_{1}$ and $w_{2})$ also contribute in $Rd_{ABC}$. For simplifying the calculations we have replaced the term $w_{0}$ by $\sqrt{1-w_{1}^{2}-w_{2}^{2}}$ in $Rd_{ABC}$. 
Our goal is to reckon the entanglement measure (three $\pi$) given in  (\ref{avp}), which demand to calculate the terms $\pi_{A}$, $\pi_{B}$ and $\pi_{c}$. First we focus on $\pi_{A}$ given in Eq. (\ref{a}), which further needs the evaluation of the terms $N_{A|BC}$, $N_{AB}$ and $N_{AC}$. To obtain $N_{A|BC}$, we take partial transpose of the matrix $Rd_{ABC}$ w.r.t. qubit A. It is given as
\begin{equation}
Rd_{ABC}^{T_{A}}=Ptr_{A}[Rd_{ABC}].
\end{equation}
Similarly for $N_{AB}$ and $N_{AC}$, first we obtain reduced density matrices $R_{AB}$ and $R_{AC}$, by tracing out the qubits C and B respectively. Further, the partial transposition is takes as
\begin{eqnarray}
R_{AB}^{T_{A}}=Ptr_{A}[Rd_{AB}],\\
R_{AC}^{T_{A}}=Ptr_{A}[Rd_{AC}].
\end{eqnarray} 
The negativity is defined as the absolute sum of negative eigenvalues of the partially transposed matrix. To calculate the negativities $N_{A|BC}$, $N_{AB}$ and $N_{AC}$ we need the eigenvalues spectrum of partially transposed matrices, which helps to simulate the negativities. The eigenvalues spectrum of matrices $Rd_{ABC}^{T_{A}}$, $Rd_{AB}^{T_{A}}$ and $Rd_{AC}^{T_{A}}$ are obtained as given below.
\begin{center}
  \begin{tabular}{ l | c}
    \hline 
    \textbf{Transposed Matrix} & \textbf{Eigenvalue Spectrum} \\  \hline \\
    $Rd_{ABC}^{T_{A}}$ & $\{0,0,0,0,x,(1-x),\sqrt{x(1-x)},\sqrt{x(1-x)}\}$  \\  \hline \\
    $Rd_{AB}^{T_{A}}$ & $\{x,y,p,p\}$  \\  \hline  \\
     $Rd_{AC}^{T_{A}}$ & $\{x,1-w_{1}^{2}-w_{2}^{2},q,q\}$  \\   \hline
  \end{tabular}
\end{center}  
with \\
\begin{center}
\begin{tabular}{ l | c  }
$x=\left(w_{1}\sin[2 \text{Dt}]-w_{2}\cos[2\text{Dt}]\right)^{2}$ &  $y=\left(w_{2}\sin[2 \text{Dt}]+w_{1}\cos[2\text{Dt}]\right)^{2}$  \\ \\
$p=\frac{1}{4}[2-2 w_{1}^{2}-2 w_{2}^{2}-\sqrt{2p_{1}}]$  & $q=\{\frac{1}{8} \left(4 \left(e w_2+f w_{1}\right){}^2-
\sqrt{2q_{1}}\right)$ 
\end{tabular}
\end{center}
where \\
$p_{1}=-4 a w_{2} w_{1}^{3}+4 a w_{2}^{3}w_{1}- (b-3)w_{1}^{4}+w_{1}^{2}\left(6(b+1) w_{2}^{2}-4\right)-(b-3)w_{2}^{4}-4 w_{2}^{2}+2,$ \\
$q_{1}=4 w_{2} w_{1}^{3} (a+10 c)+w_{1}^{4} (b+20d-13)-2 
w_{1}^{2}\left((3 b+13) w_2^2+8 (d-1)\right)+w_{2}^{4} (b-20 d-13)-8 c w_{2} w_{1} \left((d-5)w_{2}^{2}+4\right)+32 f^{2}w_{2}^{2}$ \\ \\
and
\begin{center}
\begin{tabular}{l|c|r}
$a=\sin[8Dt]$ & $b=\cos[8Dt]$ & $c=\sin[4Dt]$ \\ 
$d=\cos[4Dt]$ & $e=\sin[2Dt]$ & $f=\cos[2Dt]$.
\end{tabular}
\end{center}
\begin{figure*}
        \centering
                \includegraphics[width=0.8\textwidth]{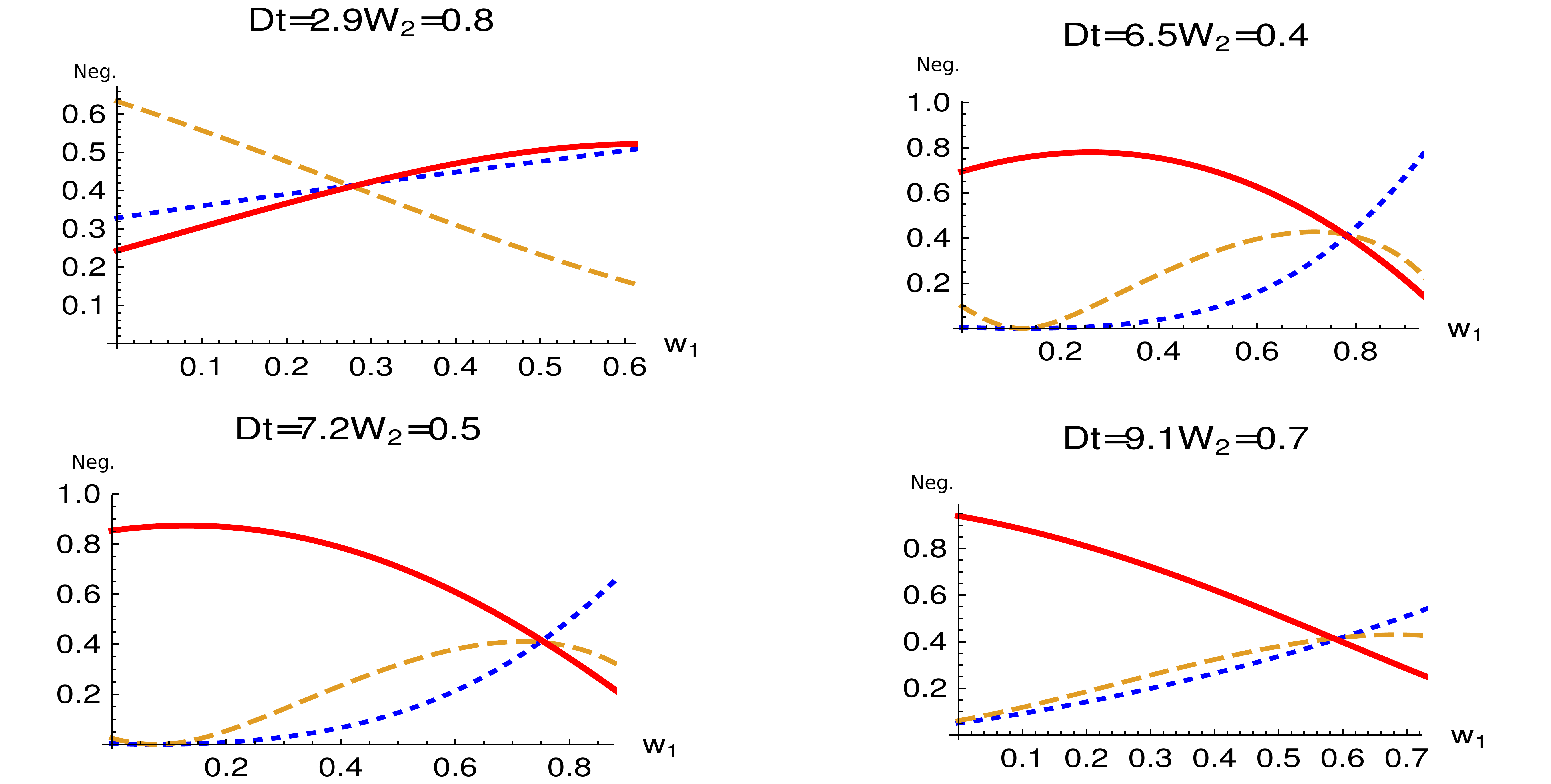}
                \caption{Plots of bipartite entanglement in generalised W states with different values of parameter $Dt$ and $w_{2}$. The quantities $N_{AB}$, $N_{AC}$ and $N_{BC}$ are represented with Dotted blue, Dashed orange and thick red graph respectively.}\label{6}
\end{figure*}
The eigenvalues spectrum depend on the parameters $w_{1}$, $w_{2}$ and $Dt$. By varying these parameters the negativities $N_{A|BC}$, $N_{AB}$ and $N_{AC}$ are obtained, which are used to calculate the term $\pi_{A}$ given in Eq. (\ref{a}). By using the above similar procedure we calculate the another negativities involved in the equations Eq. (\ref{b}) and (\ref{c}), which are further used to evaluate the terms $\pi_{B}$ and $\pi_{C}$. Finally the terms $\pi_{A}$, $\pi_{B}$ and $\pi_{C}$, lead the calculation of three $\pi$ measure given in Eq. (\ref{avp}). We here recall that, three $\pi$ measure also involve the parameters $Dt$, $w_{1}$and $w_{2}$. By varying these parameters we have to obtain the results. 

We plot the tripartite entanglement (i.e. three $\pi$) for generalized W states for different values of the parameter $Dt$ in Fig. \ref{2}. We mention that the parameter $Dt$ can be assumed as DM interaction strength with unit time of interval. In the absence of DM interaction with $Dt=0$ we find that the three $\pi$ achieve positive values for generalised W states. As the DM interaction strength increases the amplitude of three $\pi$ oscillates, because three $\pi$ involve oscillatory terms $Sin[Dt]$ and $Cos[Dt]$, which can also be observed in eigenvalues spectrum of partially transposed matrices. To make the results more clear, we plot the two dimensional plots of Fig. \ref{2} in Fig. \ref{3}. In figures of Fig. \ref{3}, we plot three $\pi$ with less range of amplitude to visualise sudden death results. 
In Fig. \ref{3}, DM interaction strength produces the periodic oscillations in three $\pi$. The entanglement falls to zero level and then rises, it does not stay at zero level for finite time. So no tripartite ESD is found in W states.

In search of ESD produced by DM interaction in generalized W states, we also study the bipartite entanglement in three partitions of tripartite system ABC. These partitions are $AB$, $AC$ and $BC$. The negativities in these partitions $N_{AB}$, $N_{AC}$ and $N_{BC}$ are plotted with different values of the set of parameters $(Dt, w_{2})$ vs. $w_{1}$ in Fig. \ref{4} and Fig. \ref{5}. The quantities $N_{AB}$, $N_{AC}$ and $N_{BC}$ are plotted by the dotted blue, dashed orange and red graphs respectively. Looking at figures of Fig. \ref{4} and \ref{5}, we find that the entanglement in partition BC is strongly present within $(0\leq Dt \leq 0.7)$. On the other hand the entanglement in partition AC is very much fragile and suffer from ESD. The entanglement in Fig. \ref{4} starts to die in partition AC at $w_{1}=0.4$ and continue till $w_{1}=0.5$. However the partition AB has initial entanglement zero, which sometimes overlap with the sudden death part of partition AC. This absence of entanglement in partitions AB and AC is filled by the entanglement present in partition BC, which maintain the total tripartite entanglement greater than zero in the system, because there is no tripartite sudden death of entanglement. 
Further in Fig. \ref{5}, we observed that both the partitions AB and AC simultaneously suffer from ESD at $w_{1}=0.45$ with $(Dt=0.6,w_{2}=0.8)$, while this absence of entanglement is filled by the entanglement present in the partition BC. We have found the threshold value of the parameter $Dt$ after which the periodic behaviour of DM interaction starts. This value is $Dt=0.8$. At this threshold value the entanglement in partition AC immediately rise and achieve the value 1. Again observing figures in Fig. \ref{5}, we found that the ESD in partition BC takes place at $w_{1}=0.6$ with $(Dt=0.8, w_{2}=0.1)$. It is  because of the periodic behaviour of DM interaction. Further the ESD zone has slight shift at $w_{1}=0.3$ with $(Dt=0.9,w_{2}=0.1)$. In the absence of entanglement in partition $BC$, the entanglement in partition $AC$ achieves higher value, which maintains the value of tripartite entanglement greater than zero in the system. On the other hand we also observe  that the entanglement in partition AB is very less fragile and accept very less fluctuations. We also study the specific class of W states which is widely used in quantum information and have symmetric properties. This state can be obtained by putting $(w_{0}=w_{1}=w_{2}=\frac{1}{\sqrt{3}})$ in Eq. (\ref{ws}). The entanglement in this state with increasing values of parameter $Dt$ are given below. \\
\begin{center}
\begin{tabular}{l*{9}{c}r}
\hline
Dt  & 0.0 & 0.1 & 0.2 & 0.3 & 0.4 & 0.5 & 0.6 & 0.7 & 0.8 \\
\hline 
Entanglement. & 0.549364 & 0.471871 & 0.281458 & 0.0819103 & 0.000557363 & 0.106773 & 0.312717 & 0.492105 & 0.547631 \\
\hline
\end{tabular}
\end{center}
The results given in the above table revels that as DM interaction strength increases, the value of entanglement decreases up to $0.000557363$. Which starts to rise immediately after $Dt=0.4$ and the state does not suffer from ESD. We plot further results in Fig. \ref{6}. In figures of Fig. \ref{6} we obtain the points where all the graphs cut each other for different values of the set of parameters $(Dt,w_{1}, w_{2})$. These values are shown in the table below.
\begin{center}
\begin{tabular}{l|c|r|c}
\hline 
\textbf{$Dt$} & \textbf{$w_{1}$} & \textbf{$w_{2}$}  & Entanglement. \\ 
\noalign{\smallskip}\hline\noalign{\smallskip}
2.9 & 2.9 & 0.8 & 0.4\\ \hline 
6.5 & 0.8 & 0.4 & 0.4 \\ \hline 
7.2 & 0.75 & 0.5 & 0.4 \\ \hline 
9.1 & 0.6 & 0.7 & 0.4 \\
\noalign{\smallskip}\hline
\end{tabular}
\end{center}
We observe that  for these values the entanglement in all the partitions AB, AC and BC achieve same amplitude as $0.4$ and no bipartite partition goes under ESD. So, for these values the entanglement is equally distributed in the partitions AB, AC and BC.
\subsection{Case 2: Initial preparation with GHZ states}
In this section, we study the dynamics of tripartite system when it is initially prepared in generalised GHZ states. We apply the procedure to obtain the reduce density matrix as adopted in case 1, after extensive calculations the reduced density matrix of the system ABC is obtained as given below .
\begin{center}
$\left[
 \begin{array}{cccccccc}
 \left(|c_0|^2+|c_1|^2\right) g_0^2 & 0 & 0 & 0 & 0 & 0 & 0 & \left(|c_0|^2+|c_1|^2\right) g_0 g_1 \\
 0 & 0 & 0 & 0 & 0 & 0 & 0 & 0 \\
 0 & 0 & 0 & 0 & 0 & 0 & 0 & 0 \\
 0 & 0 & 0 & 0 & 0 & 0 & 0 & 0 \\
 0 & 0 & 0 & 0 & 0 & 0 & 0 & 0 \\
 0 & 0 & 0 & 0 & 0 & 0 & 0 & 0 \\
 0 & 0 & 0 & 0 & 0 & 0 & 0 & 0 \\
 \left(|c_0|^2+|c_1|^2\right) g_0 g_1 & 0 & 0 & 0 & 0 & 0 & 0 & \left(|c_0|^2+|c_1|^2\right) g_1^2 \\
\end{array}\right]$
\end{center}
We find that the reduced density matrix does not involve the parameter $Dt$, which becomes zero in calculations. The matrix only involve the probability amplitudes of environmental qubit C. Further the factor $|c_{0}|^{2}+|c_{1}|^{2}$ is involved with every terms of the matrix, so by applying the normalization condition from Eq. (\ref{pa}), the probability amplitudes vanish from the reduced density matrix and it maps to initial generalized GHZ states. So, we conclude that these states are neither affected by DM interaction nor by the state of environmental qubit D. Now it is obvious that the tripartite and bipartite entanglement has not been affected in GHZ states. Hence GHZ state exhibit the robust character against DM interaction. 

\section{Conclusion}
The robustness of W and GHZ states are investigated against the Dzyaloshinskii-Moriya interaction (DM). We prepare the tripartite system in generalised W and GHZ states respectively. By taking an environmental qubit which interacts with any one of the qubit of tripartite system, we study the dynamics of entanglement in W and GHZ states against the DM interaction. In this study three $\pi$ is used as an entanglement measure. We find  that the state of environmental qubit does not contribute in the dynamics irrespective the system either prepared in W or GHZ states. It is the DM interaction which only influences the entanglement. Further we find that the tripartite entanglement does not face ESD, when system is initially prepared in W states. While on the other hand the bipartite entanglement based on negativity in W states goes under periodic ESD. This periodic behaviour takes place with threshold value of the parameter $Dt=0.8$. Tripartite entanglement in W states are more robust against DM interaction while bipartite partitions are fragile. On the other hand GHZ sates do not get affected neither by DM interaction nor by state of environmental qubit. So, we find GHZ states are more robust than W states against DM interaction. This study may be useful in quantum information processing where unwanted DM interaction takes place.


\begin{thebibliography}{99}
\bibitem{EPR1935}
Einstein, A., Podolsky, B., Rosen, N.: Can quantum-mechanical description of physical reality be considered complete? Phys. Rev. \textbf{47}, 777 (1935). 
\bibitem{Neilsen2000}
Nielsen, M. A., Chuang, I. L.: \emph{Quantum Computation and Quantum Information.} (Cambridge, U.K.: Cambridge University Press 2000).
\bibitem{sc1}
Schr{\"o}dinger, E.: Discussion of probability relations between separated systems. Mathematical Proceedings of the Cambridge Philosophical Society \textbf{31}, 555 (1935). 
\bibitem{sc2}
Schr{\"o}dinger, E.: Probability relations between separated systems. Mathematical Proceedings of the Cambridge Philosophical Society \textbf{32}, 446 (1936). 
\bibitem{bent}
Bennett, C. H., Brassard, G., Crépeau, C., Jozsa, R., Peres, A., Wootters, W. K.: Teleporting an unknown quantum state via dual classical and Einstein\textendash Podolsky\textendash Rosen channels. Phys. Rev. Lett. \textbf{70}, 1895 (1993).
\bibitem{ex}
Bouwmeester, D., Pan, J. W., Mattle, K., Eibl, M., Weinfurter, H., Zeilinger, A.: Experimental quantum teleportation. Nature \textbf{390}, 575 (1997).
\bibitem{AkEkert1991}
Ekert, A. K.: Quantum cryptography based on Bell’s theorem. Phys. Rev. Lett. \textbf{67}, 661 (1991).
\bibitem{qs}
Meyer, D.: Quantum Strategies. Phys. Rev. Lett. \textbf{82}, 1052 (1999).
\bibitem{qi}
Lugiato, L.: Quantum Imaging. J. Opt. B: Quantum Semiclass.\textbf{ 4}, 3 (2002).
\bibitem{qc}
Hillery, M., Bužek, V., Berthiaume, A.: Quantum secret sharing. Phys. Rev. A \textbf{59}, 1829 (1999). 
\bibitem{qm1}
Wiebe, N., Kapoor, A., Svorey, K.M.: Quantum algorithms for nearest-neighbor methods for supervised and unsupervised learning. arXiv:1401.2142v2 (2014).
\bibitem{qm2}
Lloyd, S., Mohseni, Rebentrost, M. P.: Quantum algorithms for supervised and unsupervised machine learning. arXiv:1307.0411v2 (2014).
\bibitem{qm3}
Yoo, S., Bang, J., Lee, C., Lee, J.: A quantum speedup in machine learning: finding an N-bit Boolean function for a classification. New Journal of Physics \textbf{16}, 103014 (2014).
\bibitem{YuEberly2004}
Yu, T., Eberly, J. H.: Finite-time disentanglement via spontaneous emission. Phys. Rev. Lett. \textbf{93}, 140404 (2004).
\bibitem{YuEberly2009}
Yu, T., Eberly, J. H.: Sudden death of entanglement. Science  \textbf{30}, 598 (2009). 

\bibitem{z1}
Qiang, Z., Xiao-Ping, Z., Qi-Jun, Z., Zhong-Zhou, R.: Entanglement dynamics of a Heisenberg chain with Dzyaloshinskii-Moriya interaction. Chin. Phys. B \textbf{18}, 3210 (2009).

\bibitem{z2}
Qiang, Z., Ping, S., Xiao-Ping, Z., Zhong-Zhou, R.:
Control of entanglement sudden death induced by Dzyaloshinskii-Moriya interaction. Chin. Phys. C \textbf{34}, 1583 (2010).

\bibitem{z3}
Qiang, Z., Qi-Jun, Z., Xiao-Ping, Z.,
Zhong-Zhou, R.: Controllable entanglement sudden birth
of Heisenberg spins. Chin. Phys. C \textbf{35}, 135 (2011).

\bibitem{k1}
Sharma, K. K., Awasthi, S. K., Pandey, S. N.: Entanglement sudden death and birth in qubit\textendash qutrit systems under Dzyaloshinskii-Moriya interaction. Quantum Inf. Process. \textbf{12}, 3437 (2013).

\bibitem{k2}
Sharma, K. K., Pandey, S. N.: Entanglement Dynamics in two parameter qubit-qutrit states under Dzyaloshinskii-Moriya interaction. Quantum Inf. Process. \textbf{13}, 2017 (2014).

\bibitem{k3}
Sharma, K. K., Pandey, S. N.: Influence of Dzyaloshinshkii-Moriya interaction on quantum correlations in two qubit Werner states and MEMS. Quantum. Info. Process. \textbf{14}, 1361 (2015).

\bibitem{k4}
Sharma K. K., Pandey, S. N.: Dzyaloshinshkii-Moriya interaction as an agent to free the bound entangled states. Quantum. Info. Process. \textbf{15}, 1539 (2016).

\bibitem{k5}
Sharma, K. K., Pandey, S. N.: Dynamics of entanglement in two parameter qubit-qutrit states with x-component of DM interaction. Commun. Theor. Phys. \textbf{65}, 278 (2016).

\bibitem{d1}
Dzyaloshinskii, I.: A thermodynamic theory of ``weak" ferromagnetism of antiferromagnetics. J. Phys. Chem. Solids \textbf{4}, 241 (1958).

\bibitem{d2}
Moriya, T.: New mechanism of anisotropic superexachange interaction. Phys. Rev. Lett. \textbf{4}, 228 (1960).

\bibitem{d3}
Moriya, T.: Anisotropic superexchange interaction and weak ferromagnetism. Phys. Rev. Lett. \textbf{120}, 91 (1960).

\bibitem{sk1}
Heinze, S., Von Bergmann, K., Menzel, M., Brede, J., Kubetzka, A.,	Wiesendanger, R., Bihlmayer, G., Blügel, S.: Spontaneous atomic scale magnetic skyrmion lattice in two dimensions. Nature Physics \textbf{7}, 713 (2011).

\bibitem{sk2}
Fert, A., Cros, V., Sampaio, J.: Skyrmions on the track. Nature Nanotechnology \textbf{8}, 152 (2013).

\bibitem{sk3}
Zhou, Y., Iacocca, E., Awad, A. A., Dumas, R. K., Zhang, F. C., Braun, H. B., Akerman, J.: Dynamically stabilized magnetic skyrmions. Nature Communications \textbf{6}, 8193 (2015).

\bibitem{sk4}
Zhang, X. C., Ezawa, M., Zhou, Y.: Magnetic skyrmion logic gates: conversion, duplication and merging of skyrmions. Sci. Rep.. \textbf{5}, 9400 (2014).

\bibitem{th1}
Zang, G. F.: Thermal entanglement and teleportation in a two-qubit Heisenberg chain with Dzyaloshinski-Moriya anisotropic antisymmetric interaction. Phys. Rev. A \textbf{75}, 034304 (2007).

\bibitem{th2}
Aky\"uz C., Aydıner  E., M\"ustecaplıoglu \"O.E.: Thermal entanglement of a two-qutrit Ising system with Dzialoshinski–Moriya interaction. Opt. Commun. \textbf{281}, 5271 (2008) .

\bibitem{gh}
Greenberger, D.M., Horne, M. A., Zeilinger, A.: Going beyond Bell's Theorem. arXiv:0712.0921 (2007).

\bibitem{w1}
D{\"u}r, W., Vidal, G., Cirac, J. I.: Three qubits can be entangled in two inequivalent ways. Phys. Rev. A \textbf{62}, 062314 (2000).
\bibitem{w2}
Fleischhauer, M., Lukin, M. D.: Quantum memory for photons: Dark-state polaritons. Phys. Rev. A \textbf{65}, 022314 (2002).

\bibitem{o1}
Zang, X. P., Yang. M., Ozaydin, F., Song, W., Cao, Z. L.:
Deterministic generation of large scale atomic W states. Opt. Express \textbf{24}, 12293 (2016).

\bibitem{o2}
Zang, X. P., Yang, M., Ozaydin, F., Song, W., Cao, Z. L.: Generating multi-atom entangled W states via light-matter interface based fusion mechanism. Sci. Rep. \textbf{5}, 16245 (2015).

\bibitem{o3}
Ozaydin, F., Altintas, A. A.: Quantum Metrology: Surpassing
the shot-noise limit with Dzyaloshinskii-Moriya interaction. Sci. Rep. \textbf{5}, 16360 (2015).

\bibitem{o4}
Ozaydina, F., Altintas A. A., Yesilyurt, C., Bugud, S., Erole, V., Quantum fisher information of bipartitions of W states. Acta. Phys. Pol. A \textbf{127}, 4 (2015).

\bibitem{o5}
Ozaydin, F., Altintas A. A., Bugu, S., Yesilyurt, C., Arik, M.: Quantum Fisher Information of Several qubits in the Superposition of a GHZ and two W states with arbitrary relative Phase. Int. J. Theor. Phys. \textbf{ 53}, 3219 (2014).

\bibitem{o6}
Yi, X. J., Huang, G. Q., Wang, J. M., Quantum fisher information of a 3-qubit state. Int. J. Theor. Phys. \textbf{51}, 3458 (2012).

\bibitem{te1}
Weinstein, Y. S.: Tripartite entanglement witnesses and entanglement sudden death. Phy. Rev. A. \textbf{79}, 012318 (2009).

\bibitem{te2}
Hu, T., Ren, H., Xue, K.: Tripartite entanglement sudden death in Yang-Baxter systems. Quantum Inf. Process \textbf{10}, 705 (2011).

\bibitem{tpi}
Vedral, V., Plenio, M. B., Rippin, M. A., Knight, P. L.: Quantifying Entanglement. Phys.
Rev. Lett. \textbf{78}, 2275 (1997).

\bibitem{mn}
Cheng Ou, Y., Fan, H.: Monogamy inequality in terms of negativity for three-qubit states. Phy. Rev. A. \textbf{75}, 062308 (2007).

\bibitem{n1}
Plenio. M. B., Virmani, S.: An introduction to entanglement measures. Quantum Inf. Comput. \textbf{7}, 1 (2007).

\bibitem{n2}
Vidal, G., Werner, R. F.: Computable measure of entanglement. Phys. Rev. A \textbf{65}, 032314 (2002).

\bibitem{n3}
Horodecki, M., Horodecki, P., Horodecki, R.: Separability of mixed states: necessary and sufficient conditions. Phys. Lett. A \textbf{223}, 1 (1996).

\bibitem{n4}
Peres, A.: Quantum Theory: Concepts and Methods (Kluwer Academic Publishers, Dordrecht, Netherlands, 1995).

\bibitem{n5}
Peres, A.: Higher order Schmidt decompositions. Phys. Lett. A \textbf{202}, 16 (1995).

\bibitem{ckw}
Coffman, V., Kundu, J., Wootters, W. K.: Distributed entanglement. Phys. Rev. A \textbf{61}, 052306 (2000).

\bibitem{proof}
Li, Q., Cui, J., Wnag, S., Long, G. L.: Study of a monogamous entanglement measure for three-qubit quantum systems. Quantum Inf. Process \textbf{15}, 2405 (2016). 
\end{thebibliography}
\end{document}